# The effect of norm-based messages on reading and understanding COVID-19 pandemic response governmental rules

Ennio Bilancini[1], Leonardo Boncinelli[2], Valerio Capraro[3], Tatiana Celadin[1], Roberto Di Paolo[1,4]

[1] IMT School of Advanced Studies, Lucca, Italy, [2] University of Florence, Italy, [3] Middlesex University London, [4] University of Alicante.

**Abstract**

The new coronavirus disease (COVID-19) threatens the lives of millions of people around the world, making it the largest health threat in recent times. Billions of people around the world are asked to adhere to strict shelter-in-place rules, finalised to slow down the spread of the virus. Appeals and messages are being used by leaders and policy-makers to promote pandemic response. Given the stakes at play, it is thus important for social scientists to explore which messages are most effective in promoting pandemic response. In fact, some papers in the last month have explored the effect of several messages on people's *intentions* to engage in pandemic response behaviour. In this paper, we make two contributions. First, we explore the effect of messages on people's *actual* engagement, and not on intentions. Specifically, our dependent variables are the level of understanding of official COVID-19 pandemic response governmental informative panels, measured through comprehension questions, and the time spent on reading these rules. Second, we test a novel set of appeals built through the theory of norms. One message targets the personal norm (what people think is the right thing to do), one targets the descriptive norm (what people think others are doing), and one targets the injunctive norm (what people think others approve or disapprove of). Our experiment is conducted online with a representative (with respect to gender, age, and location) sample of Italians. Norms are made salient using a flier. We find that norm-based fliers had no effect on comprehension and on time spent on the panels. These results suggest that norm-based interventions through fliers have very little impact on people's reading and understanding of COVID-19 pandemic response governmental rules.

**1. Introduction**

At the time in which we write (May 1, 2020), over 3 million people worldwide have been affected by the disease COVID-19, caused by the new coronavirus (SARS-CoV-2). More than 230,000 people are confirmed dead[1], and this is likely to be a severe underestimation (Burn-Murdoch, Romei, & Giles, 2020). To stop the exponential spread of the virus, dozens of countries have implemented shelter-in-place rules to the point that, at the moment, about one third of the world population is under some form of restriction (Kaplan, Frias, & McFall-Johnsen, 2020).

While medical scientists work hard to find a cure or a vaccine, the role of social and behavioural scientists is to give insights that can help align human behaviour with the recommendations of epidemiologists and public health experts (Van Bavel et al. 2020). These insights include finding efficient mechanisms to inform the population and drive behavioural changes, with the overarching goal of promoting pandemic response and minimising the potentially devastating consequences that the pandemic might cause (Van Bavel et al. 2020).

Among these mechanisms, social scientists have primarily focused on which appeals and messages promote intentions to engage in prevention behaviours (Barari et al. 2020; Capraro & Barcelo, 2020; Everett et al. 2020; Falco & Zaccagni, 2020; Heffner et al. 2020; Jordan et al. 2020; Lunn et al. 2020; Pfattheicher et al. 2020). The importance of finding efficient messages is clear, as they represent an easy and potentially scalable intervention: messages can be texted by phone, spread on social media, put inside postal boxes, and even voiced in the streets using cars equipped with a megaphone, as it happened in Italy (Provantini & Ugolini, 2020). Yet, one important limitation of these works is that they focus on *intentions* to engage in behaviours related to pandemic response, and not on *actual* engagement (Gollwitzer et al. 2020).

In this paper, we make two contributions. The first one is methodological: we develop an experimental design aimed at measuring pandemic relevant actual behaviours. To this end, compared to previous works, we consider a different dependent measure: instead of focusing directly on behaviours such as practicing physical distancing (Everett et al., 2020; Jordan et al. 2020) or wearing a face covering (Capraro & Barcelo, 2020), which are clearly hard to measure in reality, we focus on reading detailed and official information about the coronavirus. This measure is incentivised, not with money, of course, but with time. Specifically, participants in our experiment will read a series of detailed information regarding the coronavirus and then will be asked some comprehension questions. Our primary dependent measure will be the percentage of correct answers (which, as we will see, is correlated with the time spent on the panels).

Our second contribution is practical: we test a new set of messages to promote (our measure of) pandemic response. To develop this set of messages, we take a theory-driven approach. More than a century of research in social science has shown that people's decisions are affected by what people believe to be the norms in a given context (Durkheim, 1894/2017; Schwartz, 1977; Cialdini et al.,1990; Bicchieri, 2005). People tend to follow what they think other people are doing (the so-called descriptive norm), what they think other people would approve of (the injunctive norm), and what they personally think is the right thing to do (the personal norm). Consequently, in recent years behavioural scientists have started using norm-based interventions to promote desirable behaviour in economic experiments (D'Adda et al. 2017; Bicchieri & Xiao, 2009; Bilancini et al. 2020; Capraro & Rand, 2018; Capraro et al. 2019; Capraro & Vanzo, 2019; Eriksson et al. 2017; Krupka & Weber, 2009; Krupka & Weber, 2012; Kimbrough & Vostroknutov, 2016) as well as in the field (Agerström et al., 2016; Croson et al., 2010; Ferraro & Price, 2013; Frey & Meier, 2004; Goldstein et al. 2008; Allcott, 2016; Hallsworth et al. 2017; Allcott & Kessler, 2019). Moreover, several recent works have highlighted the social motives behind COVID-19 prevention behaviour (Campos-Mercade et al. 2020; Lees et al. 2020; Raihani & de-Wit, 2020). This suggests that norm-based interventions may be useful to promote pandemic response (Van Bavel et al. 2020).

---

[1] https://www.worldometers.info/coronavirus/

Having this in mind, we designed, pre-registered, and conducted a four-condition, between-subjects experiment, in which participants were shown a flier before reading a series of panels containing detailed information about how to behave in response to the coronavirus threat. Each of the three "treatment" fliers targeted a different norm; a fourth flier corresponded to the baseline. The text reported in the panels was downloaded from the website of the Italian Ministry of Health. We decided to conduct the experiment using fliers and governmental information, because of the potential scalability of such intervention: the government can send a summary of the shelter-in-place rules with a flier by text message, email, or regular mail. The experiment was conducted with a representative (with respect to age, gender and location) sample of Italians.

## 2. Method

The experiment was conducted between the 22nd and the 23rd of April, 2020. It was implemented with Qualtrics. We recruited a representative (with respect to gender, age, and location) sample of 640 Italian subjects using the online platform Lucid[2]. Participants were paid 1.25€ for a 10-minute survey. A posteriori sensitivity analysis shows that this sample size is sufficient to detect an effect size of $d = 0.28$ with significance $\alpha = 0.05$ and power of $\beta = 0.80$. Participants were randomly assigned to one of four treatments. In one treatment they were shown a flier with no explicit reference to norms (baseline), while in each of the other three treatments the flier aimed at making one of the three norms (personal, descriptive, injunctive) more salient (see Figure 1). In the Baseline (N=158), we invited participants to reflect on the current emergency situation. In the Personal Norm treatment (N=165), we invited participants to reflect on which behaviours they think are right in the current emergency situation. In the Descriptive Norm treatment (N=160), we invited participants to reflect on which behaviours they think are widespread among other people in the current emergency situation. Finally, in the Injunctive Norm treatment (N=157), we invited participants to reflect on which behaviours they think other people believe to be right in the current

---

[2] Lucid is an aggregator of survey respondents from many sources. It collects basic demographic information from all their subjects, facilitating quota sampling to match the demographic census margins for many countries (including Italy).

emergency situation.[3]

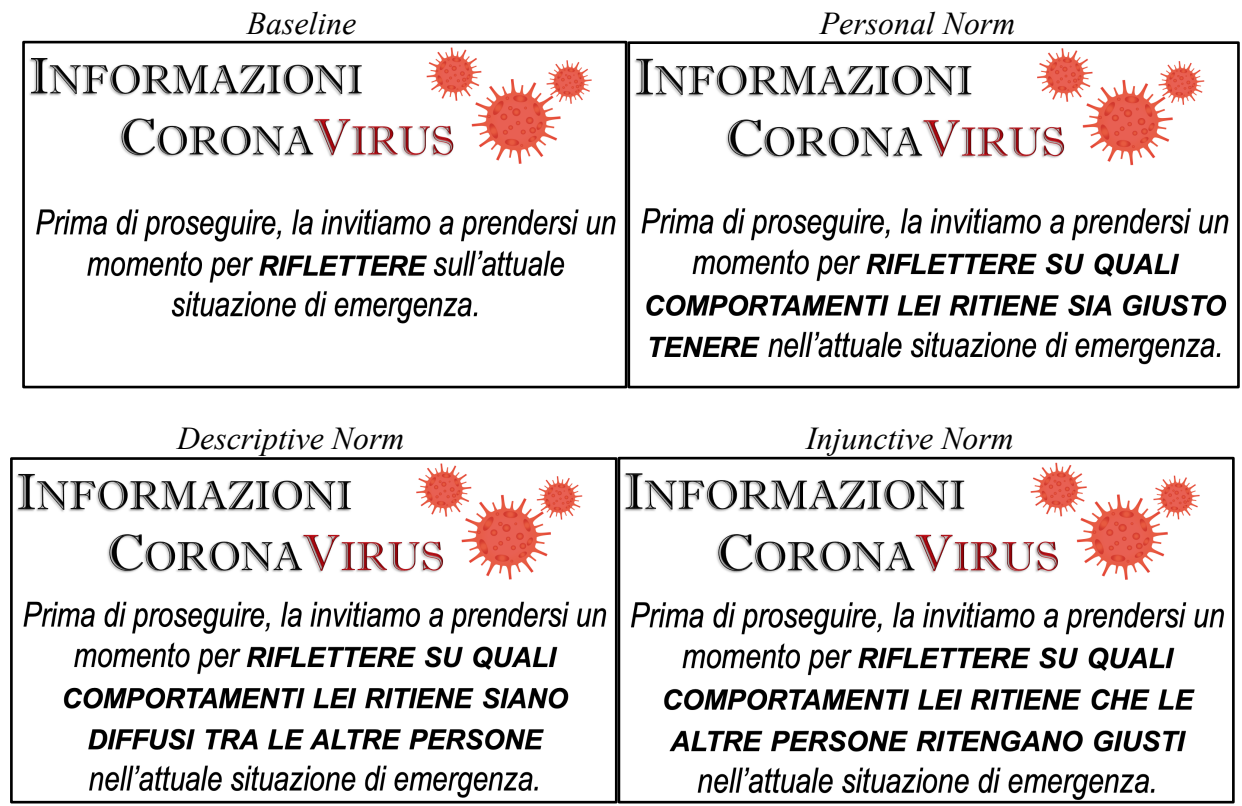

*Figure 1 - Fliers shown in each treatment. Translated from the Italian, in the Baseline, participants are invited to participants to reflect on the current emergency situation. In the Personal Norm treatment, participants are invited to reflect on which behaviours they think are right in the current emergency situation. In the Descriptive Norm treatment, participants are invited to reflect on which behaviours they think are widespread among other people in the current emergency situation. In the Injunctive Norm treatment, participants are invited to reflect on which behaviours they think other people believe to be right in the current actual emergency situation.*

After being shown the flier, participants read five informative panels about the recommended behaviours during the Covid-19 pandemic disease. The information in the panels was taken from the website of the Italian Ministry of Health[4]. A timer (invisible to the participants) allowed us to record the time that each participant spent on each panel. After each panel, participants had to answer a comprehension question about what they had just read; each question had three possible answers, one of which was correct. Correct answers were not incentivised with money, because we did not want to motivate participants to pay attention just for receiving a monetary payment. The flier was shown before each panel. However, after the first panel, the fliers were slightly different (see Appendix). The order of the panels, as well as the order of the possible answers to the questions, were fully randomised.

---

[3] An alternative way to target the descriptive norm and the injunctive norm could have been to tell participants: "90% of your neighbours think …" We opted for not using this type of nudges for two reasons: (i) to avoid deception (we do not know the location of the participants, so we cannot know what their neighbours think), and (ii) to maximise comparability with the personal norm condition (there is no way to make the personal norm salient by using a nudge of the shape "90% of…").

[4] http://www.salute.gov.it/portale/home.html

After the five panels and the corresponding five questions, participants were asked a set of demographic variables: sex, age, education, income, residence, political affiliation, general health, whether they were tested positive for COVID-19 and whether they had relatives that were tested positive for COVID-19. These measures are not explored in this paper and left for further investigation. We refer to the Appendix for full experimental instructions, both in Italian (original) and in English.

Our primary dependent variable is the percentage of correct answers given during the experiment. We investigate the effect of norm-based interventions targeting the personal norm and the two social norms (descriptive and injunctive) on this variable. Moreover, as a secondary (pre-registered) analysis, we test the differences in the distribution of times spent on the panels across treatments. The design and the analyses were pre-registered at https://aspredicted.org/th7kw.pdf.

## 3. Results

Table 1 reports the demographic characteristics of the overall sample. Our sample is representative (with respect to gender, age and location) of the Italian population. In the analysis below, we drop out one participant because she spent 25 hours on the survey, probably leaving the survey open on the computer. Given that the time spent on the panels is an important measure for our analysis, we eliminate this extreme outlier which generates a huge standard deviation in the data.

| Demographics | | Percentage in the sample | Percentage in Italian population |
|---|---|---|---|
| Gender | Female | 49.77 | 51.32 |
| | Male | 50.23 | 48.68 |
| Age | 18-24 | 8.29 | 10.21 |
| | 25-34 | 12.83 | 12.43 |
| | 35-44 | 16.59 | 15.34 |
|          | 45-54 | 19.72 | 18.43 |
|          | 55-64 | 17.68 | 15.60 |
|          | 65+   | 24.88 | 26.00 |
| Location | North | 42.72 | 45.98 |
|          | Center| 20.19 | 19.91 |
|          | South | 37.09 | 34.11 |

*Table 1 - Demographic characteristics of the sample.*

We begin by analysing the percentage of correct answers that participants give to the five questions, for each treatment. Figure 2 reports the distribution of the percentage of correct answers by treatment (left panel) and the average values of the "*percentage of correct answers*" variable split by treatment (right panel). As pre-registered, we first make an overall comparison using the Kruskal-Wallis test to identify differences in the distribution across all treatments, then we compare each treatment with the *Baseline* using the Wilcoxon rank-sum test. For the Kruskas-Wallis test we find no statistically significant difference ($X^2 = 1.272$; $p = 0.714$). Similarly, neither of the pairwise comparisons between each treatment and the *Baseline* is statistically significant (all *p*-values are larger than 0.1).

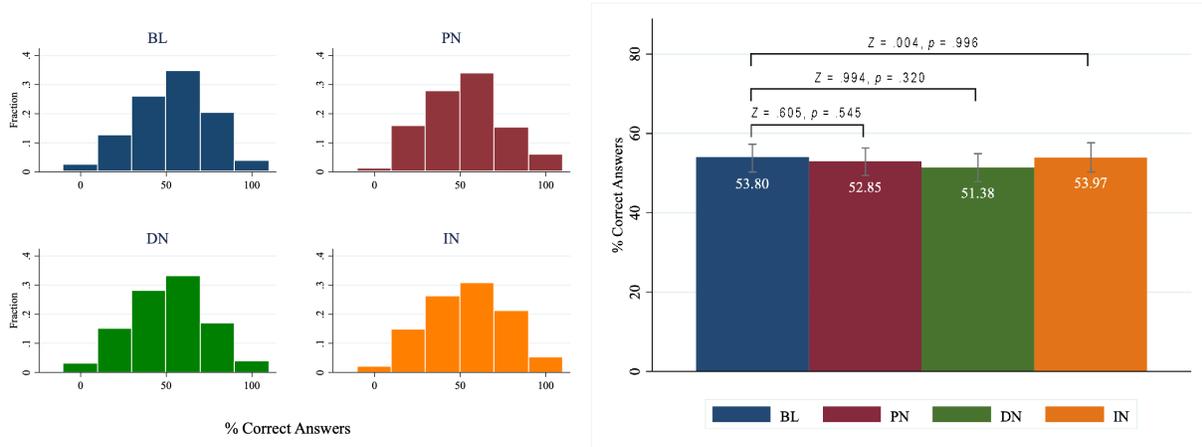

*Figure 2: Percentage of correct answers distributions, split by treatments (left chart). Average values of the "percentage of correct answers" variable by treatment (right chart). Error bars represent 95% CI.*

As (pre-registered) secondary analysis we analyse the time each participant spent on the five informative panels. The idea is to use it as a proxy for the effort that people exert to read and understand the panel. First of all, we show that the amount of time spent on the panel is positively associated with the number of correct answers, which is correlated with the understanding of the information contained in the panel itself. A linear regression predicting the number of correct answers as a function of the average time spent on the panels reports a statistically significant positive effect (coeff = 0.136, p < 0.001). Then we analyse the time spent on each panel across treatments. Figure 3 shows the average time spent on each treatment. As pre-registered, we first use the Kruskal-Wallis test to check for differences in distributions, and we then use the Wilcoxon rank-sum test to compare each treatment with the *Baseline*. For the Kruskal-Wallis test we find no statistically significant difference ($X^2$ = 4.346; $p$ = 0.226). The results of the pairwise comparisons are somewhat more equivocal: the largest effect size is found comparing the *Injunctive Norm* with the *Baseline*, but it is slightly above the conventional significance level ($Z$ = 1.932; $p$ = 0.053).

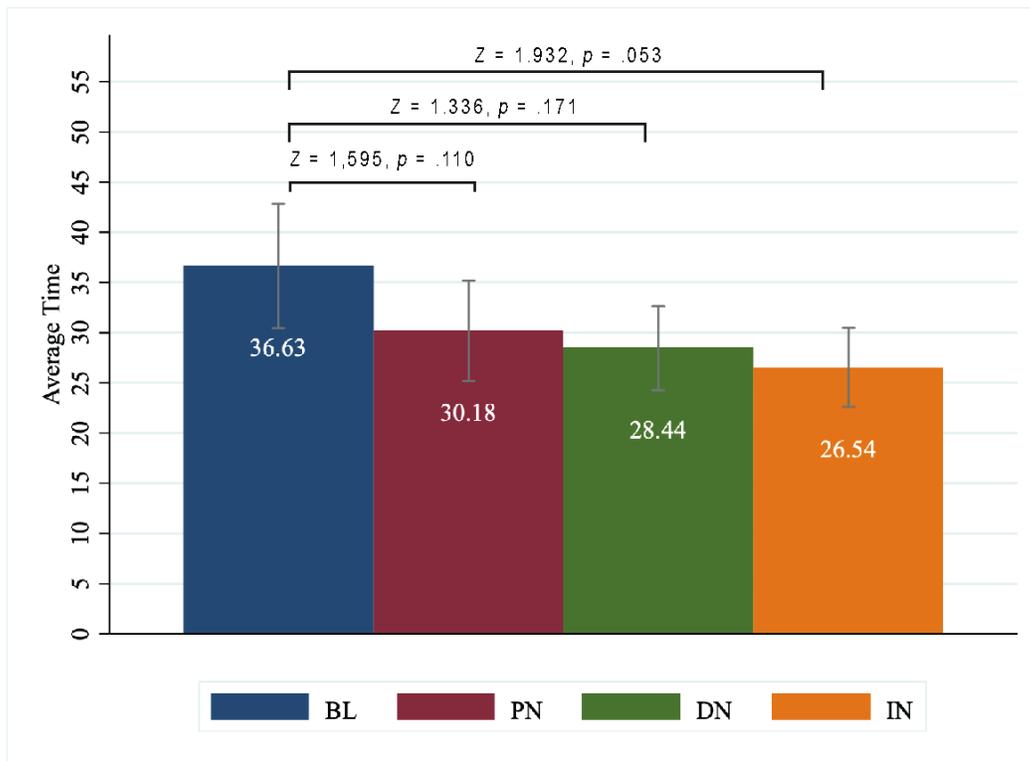

*Figure 3: Average time spent on the informative panels by treatments. Errors bars represent 95% CI.*

Finally, one may wonder if the previous results are a consequence of participants who do not read the panels, which would prevent treatments from being effective. Indeed, while analysing the data, we noticed that a substantial proportion of participants had spent far too little time on the informative panels, suggesting that they had not read them. Therefore, we tried to identify, for each informative panel, two sets of participants, those who had read it and those who had not. The left chart of Figure 4 provides evidence that the distributions of the time spent on each panel tend to be bimodal. We classified a panel as "read" by a participant if the time that the participant spent on it was larger than the minimum frequency between the two peaks in the distribution of time spent for that panel; otherwise, we classified the panel as "non-read" by the participant. Averaging over panels, the first peak corresponds to about 3 seconds spent on a panel (far too little to be able to read it), while the second peak corresponds to about 57 seconds (enough for a careful reading). Being classified as "read" turns out to be, as expected, correlated with the number of correct answers (coeff = 9.15, p < .001).

Following this categorization, we have that 50% of the participants read all the panels, while about the 20% of them did not read any of the panels. Therefore, we conducted some robustness analyses to test whether the norm-based fliers had some effects on the set of people who read all the panels. Figure 4 (right chart) shows the mean values of the *"percentage of correct answers"* split by treatments for those who effectively read all the five panels. As before, we do not find any statistically significant difference when we compare the treatments (all p-values are larger than 0.1), suggesting that previous results are robust.

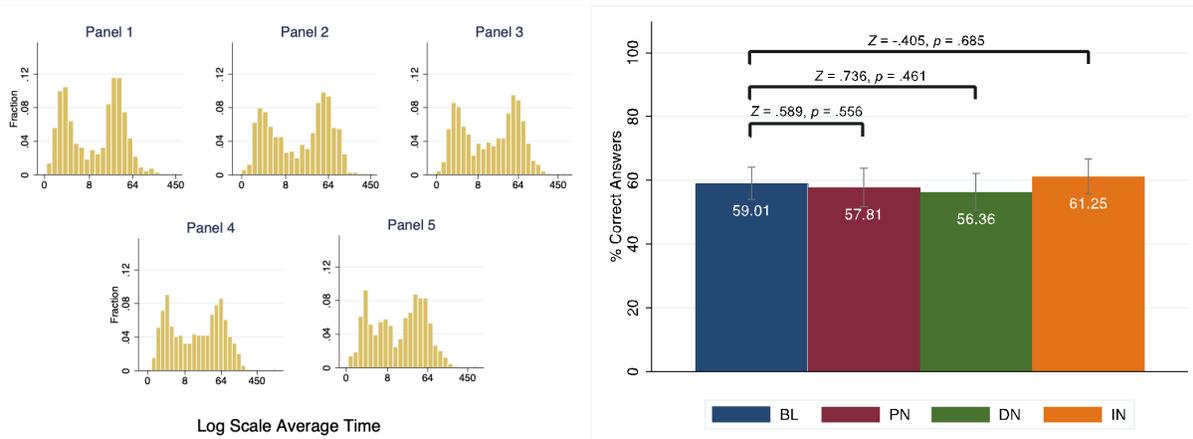

Figure 4: Distribution of the average time spent (in log scale) on each informative panel (left chart). Percentage of correct answers among participants classified as "readers" of all the five panels (right charts)

## 4. Discussion

In our experiment the norm-based interventions through fliers aimed at promoting a pandemic response had substantially no effect on reading and comprehension of the informative panels regarding behaviours recommended by the Italian Ministry of Health. The lack of a sizeable effect on comprehension was inferred from the lack of variance, across treatments, in the number of correct answers to comprehension questions administered after each informative panel. This is not an artifact of the ceiling effect as the fraction of correct answers is far below 100%, so that there was substantial room for potential improvements. The lack of a sizeable effect on reading was inferred from the lack of variance, across treatments, in the time spent in each informative panel. Actually, the distribution of time spent in each panel appears to be two-peaked, with one peak around 3 seconds and one peak around 1 minute, which suggests that participants either read the panel accurately or just skipped it, but such behaviour does not seem to be affected by our norm-based intervention.

In terms of power, a sensitivity analysis shows that our sample size was sufficient to detect a relatively small effect (d=0.28) on the primary dependent variable (percentage of correct answers). The results show very little variation on this variable across treatments and thus we believe that our sample was correctly powered to detect significant treatment effects on the primary variable. In terms of the secondary variable (time spent on the informative panels), the results are more equivocal and suggest that there might be an effect that we failed to detect because of insufficient power, such that making the norms salient has the effect that people spend less time on the panels. Paired with the fact that, when the norms are salient, people correctly answer the same number of questions as in the baseline, it is therefore possible that making the norm salient increases "efficiency of reading" (people read faster, but understand the same); an alternative explanation could be that making the norm salient has the effect that people remember that they have heard similar information in the past, and that they do not need to read the panels in detail to find the answer. Future work should test these hypotheses with a larger sample.

Our findings leave us with a warning: when actual behaviour is considered, instead of intentions, and behaviour is costly to adopt, nudging interventions should be carefully designed to be effective. Indeed, norm-based interventions may be seen as a form of nudging: they alter the choice architecture without affecting material incentives, rather relying on cognitive biases or alike (Sunstein, 2014). In light of our results, which fall possibly in the cases of ineffectiveness discussed by Sunstein (2017), we recommend future research to explore the effectiveness of stronger nudges: for instance, shocking images might be used in fliers, showing hospital wards full of sick people, or even military trucks loaded with coffins (such a kind of shocking images have already been used in a variety of situations, e.g., billboards portraying car accidents and packs of cigarettes showing smoking consequences).

Our study considered the effects of norm-based interventions on a representative sample of Italians (in terms of gender, age, and location) because we wanted to provide results about policies relying on massive and non-targeted communication. A potentially relevant direction to be explored in future research is whether the same norm-based interventions that we considered here are more effective on sub-samples with specific characteristics, such as activity on social networks or expertise of communication technologies. Given the potential heterogeneity in personal and perceived social norms, another potentially relevant direction concerns the extent to which the effects of the interventions are norm-specific (e.g., stronger effects for more extreme norms).

Moreover, we stress that the ineffectiveness of our norm-based interventions may be due to the fact that people have received so many messages and appeals to behave responsibly that no room is left for additional effects of simple nudges (such as norm-based text messages). Also, this might be especially true for online surveys and COVID-19 related studies.

We close this discussion with a methodological notice: in this paper we introduced an incentivised mechanism which is not based on money, but on effort exerted performing a task (in our case, reading the informative panel). Rather than considering this as a shortcoming, we believe it reinforces the external validity of our treatments, in that they are closer to feasible public policy interventions.

**References**

D'Adda, G., Capraro, V., & Tavoni, M. (2017). Push, don't nudge: Behavioral spillovers and policy instruments. *Economics Letters*, 154, 92–95.

Agerström, J., Carlsson, R., Nicklasson, L., & Guntell, L. (2016). Using descriptive social norms to increase charitable giving: The power of local norms. *Journal of Economic Psychology*, 52, 147–153.

Allcott, H. (2011). Social norms and energy conservation. *Journal of Public Economics*, *95*, 1082-1095.

Allcott, H., & Kessler, J. B. (2019). The welfare effects of nudges: A case study of energy use social comparisons. *American Economic Journal: Applied Economics*, *11*, 236-276.

Barari, S., et al. (2020). Evaluating COVID-19 public health messaging in Italy: Self-reported compliance and growing mental health concerns. *Available at https://www.medrxiv.org/content/10.1101/2020.03.27.20042820v2.full.pdf+html*

Bicchieri, C. (2005). *The grammar of society: The nature and dynamics of social norms*. Cambridge University Press.

Bicchieri, C., & Xiao, E. (2009). Do the right thing: but only if others do so. *Journal of Behavioral Decision Making*, 22, 191–208.

Bilancini, E., Boncinelli, L., Capraro, V., Celadin, T., & Di Paolo, R. (2020). "Do the right thing" for whom? An experiment on ingroup favouritism, group assorting and moral suasion. *Judgment and Decision Making,* 15, 182-192.

Burn-Murdoch, J., Romei, V., & Giles, C. (2020). Global coronavirus death toll could be 60% higher than reported. *Financial Times*.

Campos-Mercade, P., Meier, A. N., Schnider, F. H., Wengström, E. (2020). Prosociality predicts health behaviors during the COVID-19 pandemic. Available at *http://www.armandomeier.com/wp-content/uploads/2020/05/SocialPreferencesHealth.pdf*

Capraro, V., & Barcelo, H. (2020). The effect of messaging and gender on intentions to wear a face covering to slow down COVID-19 transmission. *Available at https://psyarxiv.com/tg7vz*.

Capraro, V., Jagfeld, G., Klein, R., Mul, M., & van de Pol, I. (2019). Increasing altruistic and cooperative be- haviour with simple moral nudges. *Scientific Reports*, 9, 11880.

Capraro, V. & Rand, D. G. (2018). Do the right thing: Experimental evidence that preferences for moral be- havior, rather than equity or efficiency per se, drive human prosociality. *Judgment and Decision Making*, 13, 99–111.

Capraro, V. & Vanzo, A. (2019). The power of moral words: Loaded language generates framing effects in the extreme dictator game. *Judgment and Decision Making*, 14, 309–317.

Cialdini, R. B., Reno, R. R., & Kallgren, C. A. (1990). A focus theory of normative conduct: recycling the concept of norms to reduce littering in public places. *Journal of personality and Social Psychology,* 58, 1015–1026.

Croson, R. T., Handy, F., and Shang, J. (2010). Gendered giving: The influence of social norms on the do- nation behavior of men and women. *International Journal of Nonprofit and Voluntary Sector Marketing*, 15, 199–213.

Durkheim, E. (2017). *Les règles de la méthode sociologique*. Flammarion..

Eriksson, K., Strimling, P., Andersson, P. A., & Lindholm, T. (2017). Costly punishment in the ultimatum game evokes moral concern, in particular when framed as payoff reduction. *Journal of Experimental Social Psychology*, 69, 59–64.

Everett, J. A., Colombatto, C., Chituc, V., Brady, W. J., & Crockett, M. (2020). The effectiveness of moral messages on public health behavioral intentions during the COVID-19 pandemic. *https://psyarxiv.com/9yqs8*

Falco, P., & Zaccagni, S. (2020). Promoting social distancing in a pandemic: Beyond the good intentions. https://doi.org/10.31219/osf.io/a2nys

Ferraro, P. J. & Price, M. K. (2013). Using nonpecuniary strategies to influence behavior: evidence from a large-scale field experiment. *Review of Economics and Statistics*, 95, 64–73.

Frey, B. S. & Meier, S. (2004). Social comparisons and pro- social behavior: Testing" conditional cooperation" in a field experiment. *American Economic Review*, 94, 1717–1722.

Goldstein, N. J., Cialdini, R. B., and Griskevicius, V. (2008). A room with a viewpoint: Using social norms to mo- tivate environmental conservation in hotels. *Journal of Consumer Research*, 35, 472–482.

Gollwitzer, A., Martel, C., Marshall, J., Höhs, J. M., & Bargh, J. A. (2020). Connecting Self-Reported Social Distancing to Real-World Behavior at the Individual and U.S. State Level. https://doi.org/10.31234/osf.io/kvnwp

Hallsworth, M., List, J. A., Metcalfe, R. D., and Vlaev, I. (2017). The behavioralist as tax collector: Using natural field experiments to enhance tax compliance. *Journal of Public Economics*, 148, 14–31.

Heffner, J., Vives, M., & FeldmanHall, O. (2020). Emotional responses to prosocial messages increase willingness to self-isolate during the COVID-19 pandemic. https://doi.org/10.31234/osf.io/qkxvb

Kimbrough, E. O., & Vostroknutov, A. (2016). Norms make preferences social. *Journal of the European Economic Association*, 14, 608–638.

Jordan, J., Yoeli, E., & Rand, D. (2020). Don't get it or don't spread it? Comparing self-interested versus prosocially framed COVID-19 prevention messaging. *https://psyarxiv.com/yuq7x/download?format=pdf*

Kaplan, J., Frias, L., & McFall-Johnsen, M. (2020). A third of the global population is on coronavirus lockdown. *Business Insider.*

Krupka, E., & Weber, R. A. (2009). The focusing and informational effects of norms on pro-social behavior. *Journal of Economic Psychology*, 30, 307–320.

Krupka, E. L., & Weber, R. A. (2013). Identifying social norms using coordination games: Why does dictator game sharing vary? *Journal of the European Economic Association*, 11, 495–524.

Lees, J. M., Cetron, J. S., Vollberg, M. C., Reggev, N., & Cikara, M. (2020). Intentions to comply with COVID-19 preventive behaviors are associated with personal beliefs, independent of perceived social norms. *https://doi.org/10.31234/osf.io/97jry*

Lunn, P. D., Timmons, S., Barjaková, M., Belton, C. A., Julienne, H., & Lavin, C. (2020). Motivating social distancing during the Covid-19 pandemic: An online experiment. *https://www.esri.ie/pubs/WP658.pdf*

Pfattheicher, S., Nockur, L., Böhm, R., Sassenrath, C., & Petersen, M. (2020). The emotional path to action: Empathy promotes physical distancing during the COVID-19 pandemic. Available at https://doi.org/10.31234/osf.io/y2cg5

Provantini, A., & Ugolini, V. (2020). Coronavirus, col megafono e sui social, sindaci in prima fila. E c'è chi requisisce anche gli asparagi. *Il Messaggero.*

Raihani, N., & de-Wit, L. (2020). Factors Associated With Concern, Behaviour & Policy Support in Response to SARS-CoV-2. *https://doi.org/10.31234/osf.io/8jpzc*

Schwartz, S. H. (1977). Normative influences on altruism. In *Advances in experimental social psychology*, volume 10, 221–279. Elsevier.

Sunstein, C. R. (2014). Nudging: a very short guide. *Journal of Consumer Policy,* 37, 583-588.

Sunstein, C. R. (2017). Nudges that fail. *Behavioural Public Policy*, 4-25.

Van Bavel, J. J., Baicker, K., Boggio, P., Capraro, V., Cichocka, A., Crockett, M., …

Willer, R. (2020). Using social and behavioural science to support COVID-19 pandemic response. *Nature Human Behaviour*.

**Appendix A.**

Here, we provide the experimental instructions of our study. Notice that we report the baseline condition, which differs from the other treatments just for the text in the flier (see Figure 1).

# BENVENUTI!

## Informazioni per i Partecipanti e Consenso

### Come saranno protette le mie informazioni?

Tutte le risposte da lei fornite saranno completamente anonime, le verrà assegnato un codice partecipante casuale che non può essere collegato in alcun modo alla sua identità personale. Se autorizza il sondaggio completandolo ed inviandolo, discuteremo/pubblicheremo i risultati in un forum accademico. In qualsiasi pubblicazione, le informazioni saranno fornite in modo tale da non poter essere identificate. Solo i membri del team di ricerca avranno accesso ai dati originali, che verranno archiviati su un computer bloccato con password. Prima che i suoi dati vengano condivisi al di fuori del gruppo di ricerca, qualsiasi informazione potenzialmente identificativa verrà rimossa. Una volta rimosse le informazioni di identificazione, i dati forniti potrebbero essere utilizzati dal team di ricerca o condivisi con altri ricercatori, sia per scopi di ricerca correlati che non correlati in futuro. I suoi dati (anonimi) possono anche essere resi disponibili in repository di dati online come Open Science Framework, che consente ad altri ricercatori e parti interessate di accedere ai dati per ulteriori analisi.

### Dichiarazione di consenso

Acconsento a partecipare a questo progetto, i cui dettagli mi sono stati spiegati e mi è stata fornita una dichiarazione scritta in un linguaggio semplice.

Capisco che la mia partecipazione a questo studio è del tutto volontaria.

Comprendo che dopo aver fatto clic sul pulsante in basso, questo modulo di consenso verrà trattenuto dal ricercatore.

Riconosco che:

(a) sono stato informato che sono libero di ritirarmi dal progetto in qualsiasi momento senza spiegazione o pregiudizio e di ritirare tutti i dati non elaborati che ho fornito;

(b) il progetto è finalizzato alla ricerca;

(c) sono stato informato che la riservatezza delle informazioni fornite sarà tutelata da eventuali requisiti legali;

(d) Qualsiasi informazione da me fornita sarà completamente anonima;

(e) Solo i membri del team di ricerca avranno accesso ai miei dati non elaborati, che verranno archiviati su un computer bloccato con password. Una volta rimosse tutte le informazioni identificabili, le mie risposte anonime possono essere condivise con altri ricercatori o rese disponibili in archivi di dati online.

**Acconsento a partecipare a questa ricerca e che le risposte che fornisco vengano trattate come sopraindicato:**

| Si, acconsento | No, NON acconsento |
| :---: | :---: |
| ○ | ○ |

**GRAZIE PER PARTECIPARE!**

Durante questo studio, leggerà cinque prospetti informativi riguardanti i comportamenti utili da tenere rispetto al Covid-19.

Tutte le informazioni sul Covid-19 qui riportate sono prese dalla pagina web del Ministero della Salute.

Successivamente ad ogni prospetto le sarà chiesto di rispondere ad una domanda.

# Informazioni CoronaVirus

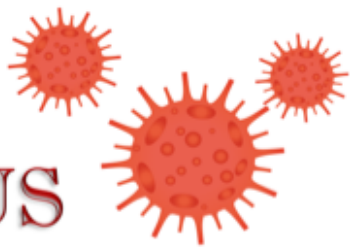

Prima di proseguire, la invitiamo a prendersi un momento per **RIFLETTERE** sull'attuale situazione di emergenza.

**Prospetto:**

Dal 26 marzo, con l'entrata in vigore del decreto-legge 25 marzo 2020, n. 19, le sanzioni sono state rese più severe e immediate. In generale, per chi viola le misure di contenimento dell'epidemia si prevede una sanzione amministrativa in denaro (da 400 a 3.000 euro). Se la violazione avviene mediante l'utilizzo di un veicolo le sanzioni possono arrivare fino a 4.000 euro. Oltre a questo, in caso di violazione delle misure di contenimento previste per pubblici esercizi, attività sportive, ludiche o di intrattenimento, attività di impresa o professionali e commerciali, può essere imposta la immediata sospensione dell'attività fino a 30 giorni. In caso di reiterazione le sanzioni pecuniarie sono raddoppiate (quindi da 800 a 6000 euro oppure 8.000 euro se commesse mediante l'utilizzo di un veicolo), mentre quella accessoria è applicata nella misura massima.

Il mancato rispetto della quarantena da parte di chi è risultato positivo al Covid-19, invece, comporta sanzioni penali: arresto da 3 a 18 mesi e pagamento di un'ammenda da 500 a 5000 euro, senza possibilità di oblazione. In ogni caso, se nel comportamento di chi commette la violazione delle misure di contenimento suddette sono riscontrati gli elementi anche di un delitto, resta la responsabilità penale per tale più grave reato. Quindi, ad esempio, rendere dichiarazioni false nelle dichiarazioni sostitutive consegnate alle forze di polizia durante i controlli resta un reato, che comporta l'immediata denuncia. Oppure violare la quarantena e, avendo contratto il virus, uscire di casa diffondendo la malattia può comportare la denuncia per gravi reati (epidemia, omicidio, lesioni), puniti con pene severe, che possono arrivare fino all'ergastolo.

**Quali sono le sanzioni che sono state introdotte con il Decreto Legge 25 marzo 2020, n. 19 laddove i soggetti vengano trovati a violare le misure di contenimento?**

- ○ Sono previste sanzioni penali e amministrative per tutti coloro che violino le misure di contenimento
- ○ Sono previste sanzioni amministrative per coloro i quali violino le misure contenitive e penali per chi risulti positivo al Covid-19
- ○ Sono previste sanzioni penali per tutti coloro che violino le misure di contenimento

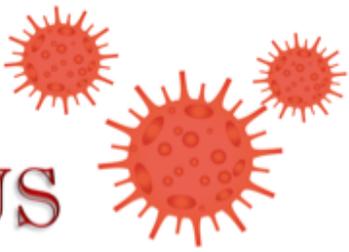

Anche in questo caso, la invitiamo a prendersi un momento per **RIFLETTERE** sull'attuale situazione di emergenza.

**Prospetto:**

Le disposizioni attualmente in vigore consentono il rientro in Italia, se si tratta di una necessità assoluta.

È quindi, per esempio, consentito il rientro dei cittadini italiani o degli stranieri residenti in Italia che si trovano all'estero in via temporanea (per turismo, affari o altro).

E' ugualmente consentito il rientro in Italia dei cittadini italiani costretti a lasciare definitivamente il Paese estero dove lavoravano o studiavano (perché, ad esempio, sono stati licenziati, hanno perso la casa, il loro corso di studi è stato definitivamente interrotto).

Una volta entrati nel territorio nazionale, gli interessati dovranno raggiungere la propria casa nel minore tempo possibile.

Le circostanze di assoluta urgenza devono essere autocertificate. Si raccomanda di preparare l'autocertificazione prima della partenza, indicando in modo specifico i motivi del rientro, in modo da rendere più rapidi i controlli.

### Per quali situazioni è possibile rientrare in Italia se si è all'estero?

○ Si può rientrare in Italia se si ha la residenza nel territorio nazionale

○ Si può rientrare in Italia previa autorizzazione della Farnesina

○ Si può rientrare in Italia solo per motivi di necessità assoluta

# Informazioni CoronaVirus 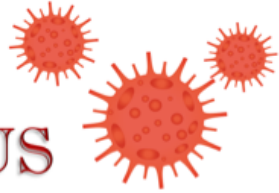

## Anche in questo caso, la invitiamo a prendersi un momento per **RIFLETTERE** sull'attuale situazione di emergenza.

**Prospetto:**

Il Governo ha emanato con il Dpcm 22 marzo 2020 nuove ulteriori misure in materia di contenimento e gestione dell'emergenza epidemiologica da COVID-19, applicabili sull'intero territorio nazionale.

Il provvedimento prevede la chiusura delle attività produttive non essenziali o strategiche.

Restano aperti alimentari, farmacie, negozi di generi di prima necessità e i servizi essenziali.

Le disposizioni producono effetto dal 23 marzo 2020 e sono efficaci fino al 3 aprile 2020.

Le stesse disposizioni si applicano, cumulativamente a quelle del Dpcm 11 marzo nonché a quelle previste dall'ordinanza del Ministro della salute del 20 marzo 2020 i cui termini di efficacia, già fissati al 25 marzo 2020, sono entrambi prorogati al 3 aprile 2020.

Tra le nuove misure è stata adottata anche l'ordinanza 22 marzo 2020, firmata congiuntamente dal Ministro della Salute e dal Ministro dell'Interno, che vieta a tutte le persone fisiche di trasferirsi o spostarsi con mezzi di trasporto pubblici o privati in comune diverso da quello in cui si trovano, salvo che per comprovate esigenze lavorative, di assoluta urgenza ovvero per motivi di salute. Ad esempio, è giustificato da ragioni di necessità spostarsi per fare la spesa, per acquistare giornali, per andare in farmacia, o comunque per acquistare beni necessari per la vita quotidiana. Inoltre è giustificata ogni uscita dal domicilio per l'attività sportiva o motoria all'aperto. In ogni caso, tutti gli spostamenti sono soggetti al divieto generale di assembramento, e quindi dell'obbligo di rispettare la distanza di sicurezza minima di 1 metro fra le persone.

**Quali misure essenziali sono state introdotte con il DPCM 22 Marzo 2020?**

○ Non è possibile uscire dalla propria abitazione, salvo che per fare jogging nei pressi della propria abitazione

○ Non è possibile spostarsi con i mezzi di trasporto pubblico in comuni diversi dal proprio, anche se per esigenze lavorative o di salute

○ Non è possibile rimanere a lungo affacciati al balcone in compagnia dei vicini

# Informazioni CoronaVirus

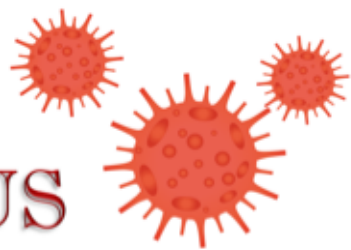

Anche in questo caso, la invitiamo a prendersi un momento per **RIFLETTERE** sull'attuale situazione di emergenza.

## Prospetto:

Per mantenerti protetto, mantieniti informato sulla diffusione della pandemia, disponibile sul sito dell'OMS e sul sito del ministero e adotta le seguenti misure di protezione personale:

· Restare a casa, uscire di casa solo per esigenze lavorative, motivi di salute e necessità
· Lavarsi spesso le mani;
· Evitare il contatto ravvicinato con persone che soffrono di infezioni respiratorie acute;
· Evitare abbracci e strette di mano;
· Mantenimento, nei contatti sociali, di una distanza interpersonale di almeno un metro;
· Igiene respiratoria (starnutire e/o tossire in un fazzoletto evitando il contatto delle mani con le secrezioni respiratorie);
· Evitare l'uso promiscuo di bottiglie e bicchieri;
· Non toccarsi occhi, naso e bocca con le mani;
· Coprirsi bocca e naso se si starnutisce o tossisce;
· Non prendere farmaci antivirali e antibiotici, a meno che siano prescritti dal medico;
· Pulire le superfici con disinfettanti a base di cloro o alcol;
· Usare la mascherina solo se si sospetta di essere malati o se si presta assistenza a persone malate.

Se presenti febbre, tosse o difficoltà respiratorie e sospetti di essere stato in stretto contatto con una persona affetta da malattia respiratoria Covid-19: rimani in casa, non recarti al pronto soccorso o presso gli studi medici ma chiama al telefono il tuo medico di famiglia, il tuo pediatra o la guardia medica. Oppure chiama il numero verde regionale. Utilizza i numeri di emergenza 112/118 soltanto se strettamente necessario.

**Gli antibiotici possono essere utili per prevenire l'infezione da Covid-19?**

- ○ Si, ma bisogna assumerli solo se prescritti dal medico
- ○ No, gli antibiotici non sono efficaci contro i virus, ma funzionano solo contro le infezioni batteriche
- ○ Si, gli antibiotici sono efficaci per prevenire l'infezione da Covid-19

# INFORMAZIONI CORONAVIRUS

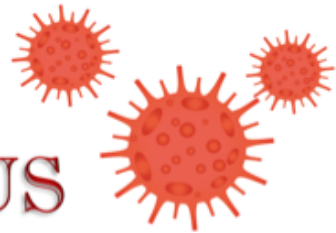

*Anche in questo caso, la invitiamo a prendersi un momento per **RIFLETTERE** sull'attuale situazione di emergenza.*

**Prospetto:**

L'attuale diffusione del Covid-19 è il risultato della trasmissione da uomo a uomo. Ad oggi, non ci sono prove che gli animali da compagnia possano diffondere il virus. Pertanto, non vi è alcuna giustificazione nell'adottare misure contro gli animali da compagnia che possano comprometterne il benessere.

Tuttavia, dal momento che gli animali e l'uomo possono talvolta condividere alcune malattie (note come malattie zoonotiche), è necessario sempre e non solo per il timore del Covid-19, che vengano adottate sempre le normali misure igieniche raccomandate da medici e veterinari per evitare la diffusione delle malattie.

Il Ministero della Salute, in accordo con quanto espresso da autorevoli Organismi internazionali, raccomanda il rispetto delle più elementari norme igieniche quali lavarsi le mani prima e dopo essere stati a contatto o aver toccato gli animali, il loro cibo o le provviste, evitare di baciarli, farsi leccare o condividere il cibo.

Al ritorno dalla passeggiata, pulire sempre le zampe evitando prodotti aggressivi e quelli a base alcolica che possono indurre fenomeni irritativi.

Piccoli accorgimenti che ci permettono di ridurre al minimo il rischio di introdurre in casa, al termine di una passeggiata, patogeni che potrebbero diffondersi negli spazi comuni.

Nelle abitazioni in cui ci sono soggetti affetti o sottoposti a cure mediche per Covid-19 si devono evitare, per quanto possibile, i contatti ravvicinati con i propri animali così come si fa per gli altri conviventi e fare in modo che se ne occupi un altro familiare.

La presenza di un animale in casa può considerarsi una grande opportunità per tutta la famiglia, sia da un punto di vista educativo che sociale.

**Gli animali da compagnia possono trasmettere il Covid-19?**

○ Gli animali da compagnia possono essere suscettibili al Covid-19, ma non c'è evidenza che siano vettori del visus

○ Gli animali da compagnia non sono suscettibili al Covid-19, e quindi non sono vettori del visus

○ Gli animali da compagnia possono essere suscettibili al Covid-19 e c'è evidenza che siano vettori del visus

Età:

[ ]

Genere:

○ Uomo

○ Donna

Indichi il livello di istruzione più alto che ha conseguito:

- ○ Scuola primaria
- ○ Scuola secondaria di primo grado
- ○ Scuola secondaria di secondo grado
- ○ Laurea di primo livello
- ○ Laurea di secondo livello
- ○ Dottorato di ricerca
- ○ Altro

Luogo di residenza

[ ]

Dov'è domiciliato attualmente? Indichi comune (Provincia)

[ ]

Indichi la sua professione:

[ ]

Tenedo in considerazione le sue condizioni di salute generali, come valuterebbe la sua salute?

| Molto cattiva | Cattiva | Né cattiva né buona | Buona | Molto buona |
|---|---|---|---|---|
| ○ | ○ | ○ | ○ | ○ |

Controlli la seguente lista di patologie.
Problemi cardiovascolari, diabete, epatite B, problemi polmonari cronici, problemi renali cronici, tumore.
Quante delle seguenti patologie attualmente ha:

| 0 | 1 | 2 | 3 | 4 | 5 o più |
|---|---|---|---|---|---------|
| ○ | ○ | ○ | ○ | ○ | ○ |

Ha avuto sintomi riconducibili al Covid-19:

○ Si

○ No

○ Preferisco non rispondere

È risultato positivo al Covid-19:

○ Si

○ No

○ Preferisco non rispondere

Ha conoscenti stretti risultati positivi al Covid 19:

○ Si

○ No

○ Preferisco non rispondere

Indichi il livello di reddito netto del 2019. Consideri tutte le forme di guadagno:

○ Fino a 15.000

○ 15.001 - 28.000

○ 28.001 - 55.000

○ 55.001 - 75.000

○ Oltre 75.001

A che partito politico è più affine?

[                                    ]

Grazie per la sua partecipazione!

Prosegua per terminare lo studio.

**Appendix B.**
Here, we provide the English translation of the experimental instructions of our study.

# WELCOME!
## Information for Participants and Consent

**How will personal information be protected?**
All the answers that you provide will be completely anonymous, you will be assigned a random participant code that cannot be linked in any way to your personal identity. If you authorize the survey by completing and submitting it, we will discuss/publish the results in an academic outlet. In any publication, the information will be provided in such a way that you cannot be identified. Only members of the research team will have access to the original data set, which will be stored on a password-locked computer. Before your data is shared outside the research team, any potentially identifying information will be removed; the data provided could be used by the research team or shared with other researchers, both for related and unrelated research purposes in the future. Your (anonymous) data may also be made available in online data repositories such as the Open Science Framework, which allows other researchers and stakeholders to access the data for further analysis.

**Declaration of consent**
I agree to participate in this project, the details of which have been explained to me and a written statement have been provided in plain language.
I understand that after clicking on the button below, this consent form will be retained by the researcher.
I recognize that:
  a) I have been informed that I am free to withdraw from the project at any time without explanation or prejudice and to withdraw all the raw data that I have provided;
  b) The project is aimed at research;
  c) I have been informed that the confidentiality of the information provided will be protected from any legal requirements;
  d) Any information I provide will be completely anonymous;
  e) Only members of the research team will have access to my raw data, which will be stored on a computer locked with a password. Once all identifiable information has been removed, my anonymous responses can be shared with other researchers or made available in online data stores.

**I agree to participate in this research and that the answers I provide are treated as indicated above:**
Agree - Disagree

*Next page*

**THANK YOU FOR PARTICIPATING!**
During this study, you will read five panels regarding the right behaviors to be followed due to Covid-19.
All information on Covid-19 reported here is taken from the web page of the Ministry of Health.
After each prospect you will be asked to answer a question.

*Next page*
*Subjects will randomly be assigned to one of the following treatments.*
*Baseline:*
**Coronavirus information**
Before continuing, we invite you to take a moment to reflect upon the current emergency situation.
*Personal Norm:*
**Coronavirus information**
Before continuing, we invite you to take a moment to reflect upon what behaviors you think are right in the current emergency situation.
*Descriptive Norm:*
**Coronavirus information**
Before continuing, we invite you to take a moment to reflect upon what behaviors you think are widespread among other people in the current emergency situation.
*Injunctive norm:*
**Coronavirus information**
Before continuing, we invite you to take a moment to reflect upon what behaviors you think other people believe to be right in the current emergency situation.

*Next page (Notice that, hereafter, we insert the baseline condition that differs from the other treatments just for the text in the flier.)*

**Panel:**
Since the 26th of March, with the entry into force of the decree-law 25 March 2020, n. 19, the sanctions have been made more severe and immediate. In general, for those who violate the restrictions to contain the epidemic, an administrative sanction in cash is foreseen (from 400 to 3,000 euros). If the violation occurs through the use of a vehicle, the penalties can reach up to 4,000 euros. In addition to this, in case of violation of the containment measures provided for public exercises, sports, leisure or entertainment activities, business or professional and commercial activities, the immediate suspension of the activity up to 30 days may be imposed. In case of reiteration, the fines are doubled (therefore from 800 to 6,000 euros or 8,000 euros if committed through the use of a vehicle), while the ancillary one is applied to the maximum extent.

Failure to comply with the quarantine by those who tested positive for Covid-19, on the other hand, entails criminal penalties: detention from 3 to 18 months and payment of a fine from 500 to 5,000 euros, without the possibility of "oblation".[5] In any case, if the elements of a crime are also found in the behavior of those who violate the aforementioned containment measures, the criminal responsibility for this more serious crime remains. So, for example, making false statements in substitute statements delivered to the police force during checks remains a crime, which requires immediate reporting. Or violate the quarantine and, having contracted the virus, leaving the house spreading the disease can lead to reporting for serious crimes (epidemic, murder, injury), punished with severe penalties, which can go as far as life in prison.

*Next Page*

**What are the sanctions that have been introduced with Law Decree 25 March 2020, n. 19 where the subjects are found to violate the containment measures?**
- Administrative penalties are provided for those who violate containment and criminal penalties for those who test positive for Covid-19
- Criminal penalties are provided for all those who violate containment measures
- Criminal and administrative penalties are provided for all those who violate containment measures

*Next Page*

**Coronavirus information**
Again, we invite you to take a moment to reflect on the current emergency situation

*Next Page*

**Panel:**
The provisions currently in force allow the return to Italy, if it is an absolute necessity.
Therefore, for example, the return of Italian citizens or foreigners residing in Italy who are abroad temporarily (for tourism, business or otherwise) is allowed.
Italian citizens forced to permanently leave the foreign country where they worked or studied are also allowed to return to Italy (because, for example, they were fired, they lost their home, their course of study was definitively interrupted).
Once they enter the national territory, the interested parties must reach their home in the shortest possible time.

---

[5] Oblation, in the Italian code of law, represents the possibility to pay a certain sum of money to extinguish the crime.

Circumstances of absolute urgency must be self-certified. It is recommended to prepare the self-certification before departure, indicating specifically the reasons for the return, in a way to speed up the checks.

*Next Page*

**For which situations can you return to Italy if you are abroad?**
- You can go back to Italy only for reasons of absolute necessity
- You can go back to Italy if you have your residence in the national territory
- You can go back to Italy with the authorization of the Farnesina

*Next Page*

**Coronavirus information**
Again, we invite you to take a moment to reflect on the current emergency situation

*Next Page*

**Panel:**
With the Ministerial Decree of 22 March 2020, the Government issued new additional measures regarding the containment and management of the epidemiological emergency from COVID-19, applicable throughout the country.
The provision provides for the closure of non-essential or strategic production activities. Food, pharmacies, necessities shops and essential services remain open.
The provisions take effect from March 23, 2020 and are effective until April 3, 2020.
The same provisions apply, cumulatively to those of the Prime Ministerial Decree of 11 March as well as to those provided for by the ordinance of the Minister of Health of 20 March 2020 whose terms of effectiveness, already set for 25 March 2020, are both extended to 3 April 2020.
Among the new measures, the ordinance of 22 March 2020 was also adopted, signed jointly by the Minister of Health and the Minister of the Interior, which prohibits all persons from moving with public or private means of transport in a municipality other than the one in which they are located, except for proven essential work, absolute urgency or health reasons. For example, it is justified by reasons of necessity to move to shop, to buy newspapers, to go to the pharmacy, or in any case to buy goods necessary for daily life. Furthermore, every exit from the home for outdoor sports or motor activities is justified. In any case, all movements are subject to the general assembly ban, and therefore the obligation to respect the minimum safety distance of 1 meter between people.

*Next Page*

**What essential measures have been introduced with the Prime Ministerial Decree of March 25, 2020?**
- It is not possible to leave your home, except for jogging close to your home
- It is not possible to travel by public transport in municipalities other than your own, even if for essential work or health reasons
- It is not possible to stay on the balcony for a long time in the company of neighbors

*Next Page*

**Coronavirus information**
Again, we invite you to take a moment to reflect on the current emergency situation

*Next Page*

**Panel:**
To keep yourself protected, stay informed on the spread of the pandemic, available on the WHO website and the ministry website, and take the following personal protection measures:
- Stay at home, leave it only for essential work, health reasons and necessities (see containment measures)
- Wash your hands often;
- Avoid close contact with people suffering from acute respiratory infections;
- Avoid hugs and handshakes;
- Maintaining, in social contacts, the interpersonal distance of at least one meter;
- Respiratory hygiene (sneeze and/or cough in a handkerchief, avoid contact of the hands with respiratory secretions);
- Avoid the promiscuous use of bottles and glasses;
- Do not touch your eyes, nose and mouth with your hands;
- Cover your mouth and nose if you sneeze or cough;
- Do not take antiviral drugs and antibiotics unless prescribed by your doctor;
- Clean the surfaces with chlorine or alcohol based disinfectants;
- Use the mask only if you suspect that you are sick or if you are caring for sick people.

If you have a fever, cough or breathing difficulties and suspect that you have been in close contact with a person suffering from Covid-19 respiratory disease: stay at home, do not go to the emergency room or doctor's office but call your family doctor on the phone, your paediatrician or the medical guard. Or call the regional toll-free number. Use the emergency numbers 112/118 only if strictly necessary.

*Next Page*

**Can Antibiotics Help Prevent Covid-19 Infection?**

- No, antibiotics are not effective against viruses, but only work against bacterial infections
- Yes, antibiotics are effective for preventing Covid-19 infection
- Yes, but you should take them only if prescribed by your doctor

*Next Page*

**Coronavirus information**
Again, we invite you to take a moment to reflect on the current emergency situation

*Next Page*

**Panel:**
The current spread of Covid-19 is the result of human-to-human transmission. To date, there is no evidence that pets can spread the virus. Therefore, there is no justification for taking measures against pets that could compromise their well-being.
However, since animals and humans can sometimes share certain diseases (known as zoonotic diseases), it is always necessary and not only for fear of Covid-19, that the normal hygiene measures recommended by doctors and veterinarians are always adopted to avoid the spread of disease.
The Ministry of Health, following what has been expressed by authoritative international bodies, recommends compliance with the most basic hygiene rules such as washing hands before and after being in contact or having touched animals, their food or supplies, avoid kissing them, get licked or share food.
Upon returning from a walk, always clean their legs avoiding aggressive products and those based on alcohol which can induce irritative phenomena.
Small tricks that allow us to minimize the risk of introducing into the home, at the end of a walk, pathogens that could spread in the common areas.
In homes where there are subjects affected or undergoing medical treatment for Covid-19, close contact with their animals should be avoided as far as possible, as is the case with other cohabitants and ensure that another one is taken care of relatives.
The presence of an animal in the house can be considered a great opportunity for the whole family, both from an educational and social point of view.

*Next Page*

**Can pets transmit Covid-19?**
- Pets may be susceptible to Covid-19, but there is no evidence that they are vectors of the virus
- Pets may be susceptible to Covid-19 and there is evidence that they are vectors of the virus

- Pets are not susceptible to Covid-19, and therefore are not vectors of the virus

*Next Page*

**Age**:
**Gender**: Man-Woman
**Indicate the highest level of education you have achieved**:
   - Primary school
   - First grade secondary school
   - High school
   - Bachelor's degree
   - Master's degree
   - PhD
   - Other

**Place of residence:**
**Where are you currently domiciled? Indicate municipality (Province)**:
**Indicate your profession:**
**Taking your general health condition into consideration, how would you rate your health?**
Very bad/Bad/Neither bad nor good/Good/Very good
**Check the following list of pathologies.**
**Cardiovascular problems, diabetes, hepatitis B, chronic lung problems, chronic kidney problems, cancer.**
**How many of the following conditions do you currently have:**
0 1 2 3 4 5 or more
**Have you had symptoms related to Covid-19?** Yes-No-Rather not to answer
**Hve you tested positive for Covid-19?** Yes-No-Rather not to answer
**Do you have close acquaintances who were tested positive for Covid 19:** Yes-No-Rather not to answer
**Indicate the level of net income in 2019. Consider all forms of income:**
   - Up to 15,000
   - 15.001 - 28.000
   - 28,001 - 55,000
   - 55.001 - 75.000
   - Over 75,001
**Which political party are you closer to?**

*Next Page*

Thanks for your participation!

Continue to finish the study.